\documentclass[a4paper, 10pt, oneside, twocolumn, 1p, number, sort&compress ]{elsarticle} 
\usepackage{hyperref}
 
\newcommand{\oo}{O$_2$}
\newcommand{\oom}{O$_2^{\mbox{-}}$}

\newcommand{\sqt}{c(2$\times$2)}
\newcommand{\sqtt}{c(4$\times$4)}
\newcommand{\ct}{\sqt}
\newcommand{\cf}{\sqtt}
\newcommand{\tpps}{$2p\mbox{-}\pi^\ast$}
\newcommand{\tppsper}{$2p\mbox{-}\pi_\perp^\ast$}
\newcommand{\tppspar}{$2p\mbox{-}\pi_\parallel^\ast$}

\newcommand{\srosurf}{Sr$_3$Ru$_2$O$_7$(001)}
\newcommand{\crosurf}{Ca$_3$Ru$_2$O$_7$(001)}
\newcommand{\cro}{Ca$_3$Ru$_2$O$_7$}
\newcommand{\sro}{Sr$_3$Ru$_2$O$_7$}
\newcommand{\hse}{HSE06}
\newcommand{\opt}{optB86b}
\newcommand{\gw}{G$_{0}$W$_{0}$}
\newcommand{\ea}{E_{\mathrm{ads}}}

\title{Adsorption of a superoxo \oom~species on the pure and Ca-doped \srosurf~surface}
\author[iap,cms]{Wernfried Mayr-Schm\"olzer}
\ead{wms@cms.tuwien.ac.at}
\author[iap,cms]{Florian Mittendorfer\corref{cor1}}
\ead{fmi@cms.tuwien.ac.at}
\author[iap,cms]{Josef Redinger}
\ead{jr@cms.tuwien.ac.at}
\cortext[cor1]{Corresponding author}
\address[iap]{Institute of Applied Physics, Vienna University of Technology, Wiedner Hauptstr. 8-10, 1040 Vienna, Austria}
\address[cms]{Center for Computational Materials Science, Vienna University of Technology, Wiedner Hauptstr. 8-10, 1040 Vienna, Austria}

\begin{document}

\begin{abstract}
Only recently, the activation of oxygen molecules on clean defect-free transition metal oxide surfaces has been reported, for example on the CaO-terminated surface of the Ruddelsden-Popper perovskite \crosurf. In this work we show that oxygen molecules adsorb as an activated superoxo species on a clean SrO-terminated surface of \srosurf.  At all coverages, the electrons activating the molecule originate from the subsurface RuO$_2$ layer. At low coverages, the presence of a  Ca dopant in the terminating SrO layer slightly increases the adsorption energy. At high coverage, DFT predicts a flat potential energy surface  and a preferred adsorption of the \oom\ near surface cations. Advanced many-electron calculations (RPA) predict adsorption energies of $-0.99$\,eV and $-0.49$\,eV per \oom~molecule for low and high coverages, 
respectively, and a preference for forming line-like structures in the latter case. 
\end{abstract}


\maketitle

\section{Introduction}
In recent years, the chemistry of ternary transition metal oxides has garnered increased interest due to their vast potential for applications, ranging from fuel cells to electrocatalysis. In addition, the electronic, magnetic and chemical properties of  perovskites of the Ruddelsden-Popper series are often tuneable by doping, applying electric or magnetic fields, or mechanical stress. 

Much research has been done regarding the oxygen transport and reduction properties on various perovskite materials \cite{Royer2006,Suntivich2011,May2012,Rincon2014}, but the activation of molecular oxygen is less well understood. In particular, the ability to dissociation  oxygen molecules plays a key role in current and future applications like solid oxide fuel cells (SOFC). Recently, 
theoretical and experimental studies have shown that the pristine SrO terminated surface of SrTiO$_3$ does not dissociate adsorbed \oo~molecules. Only in the presence of vacancies the surface activity is increased and adsorption of a charged superoxo \oom~species was predicted \cite{Staykov2015a}. The defect-free LaO-terminated (001) surface of another Ruddelsden-Popper perovskite, La$_2$NiO$_4$, already shows activation of the \oo~molecule for the stoichiometric surface, as charge is transferred from the surface La atoms to the adsorbed oxygen \cite{Akbay2016}. Recently, a combined theoretical and experimental study of the (001) surface of \cro~also predicts the adsorption of oxygen as a superoxo species, but in that case a different charge transfer mechanism was identified. In contrast to La$_2$NiO$_4$, charge is not transferred from the A-site cation, but from the conduction electrons close to the Fermi level \cite{Halwidl2018}.

In the present paper, we present a  theoretical study of the adsorption of \oo~molecules on the (001) surface of pure and Ca-doped \sro. \sro, a Ruddelsden-Popper perovskite structurally closely related to the aforementioned \cro, shows the same SrO surface termination after cleaving as the related binary SrO, but at a larger lattice constant \cite{Stoger2014a}. 

\section{Computational Methods}
All calculations were performed with the Vienna \textit{Ab-Initio} Simulation Package (VASP) using the projector-augmented plane wave method \cite{Kresse1999}. The electronic interactions were described within the framework of Density Functional Theory (DFT) using the spin-polarized van-der-Waals corrected \opt\ \cite{Klimes2010,Klimes2011} 
exchange-correlation functional. For the slab calculations, dense $\Gamma$-centered Monkhorst-Pack grids of
6$\times$6$\times$1 $\vec{k}$-points were used to sample the Brillouin zone and the energy cutoff for the plane waves was set to 400\,eV. 
Hybrid functional calculations were performed using the \hse\ functional \cite{Krukau2006}. 
The RPA total energy calculations are based on the Adiabatic Connection Fluctuation-Dissipation Theory (ACFDT) and were performed for the \opt-optimized structures using a modern, low-scaling algorithm \cite{Kaltak2014a}, starting from initial PBE orbitals. 
All calculations did not take zero point effect into account.

The surface structure was modeled using double-layer slabs of \sro, separated by 24\,\AA\ of vacuum. For the low coverage case a \sqtt~unit cell was used, yielding a minimum coverage of 1/16 after adsorption of a single \oo~molecule. The high coverage case was studied by a \sqt~unit cell with up to four adsorbed \oo~molecules.  The $\vec{k}$-point grids were adjusted according to the size of the reciprocal cell: for the low coverage \sqtt~cell a 3$\times$3$\times$1 and for the smaller \sqt~cell a 6$\times$6$\times$1 $\Gamma$-centered  $\vec{k}$-point grid was used, respectively. Due to the computational cost involved in the hybrid (\hse) and many-electron approaches (\gw/RPA) only simulations of the \sqt~model with a reduced 4$\times$4$\times$1 (\hse) and 3$\times$3$\times$1 (RPA)  $\Gamma$-centered $\vec{k}$-point grid,  were viable. 

\section{Results}
\subsection{Pristine Surface}
\begin{figure}[htbp]
\begin{center}
\includegraphics[trim= 0cm 0cm 0cm 0cm, clip, width=0.45\textwidth] {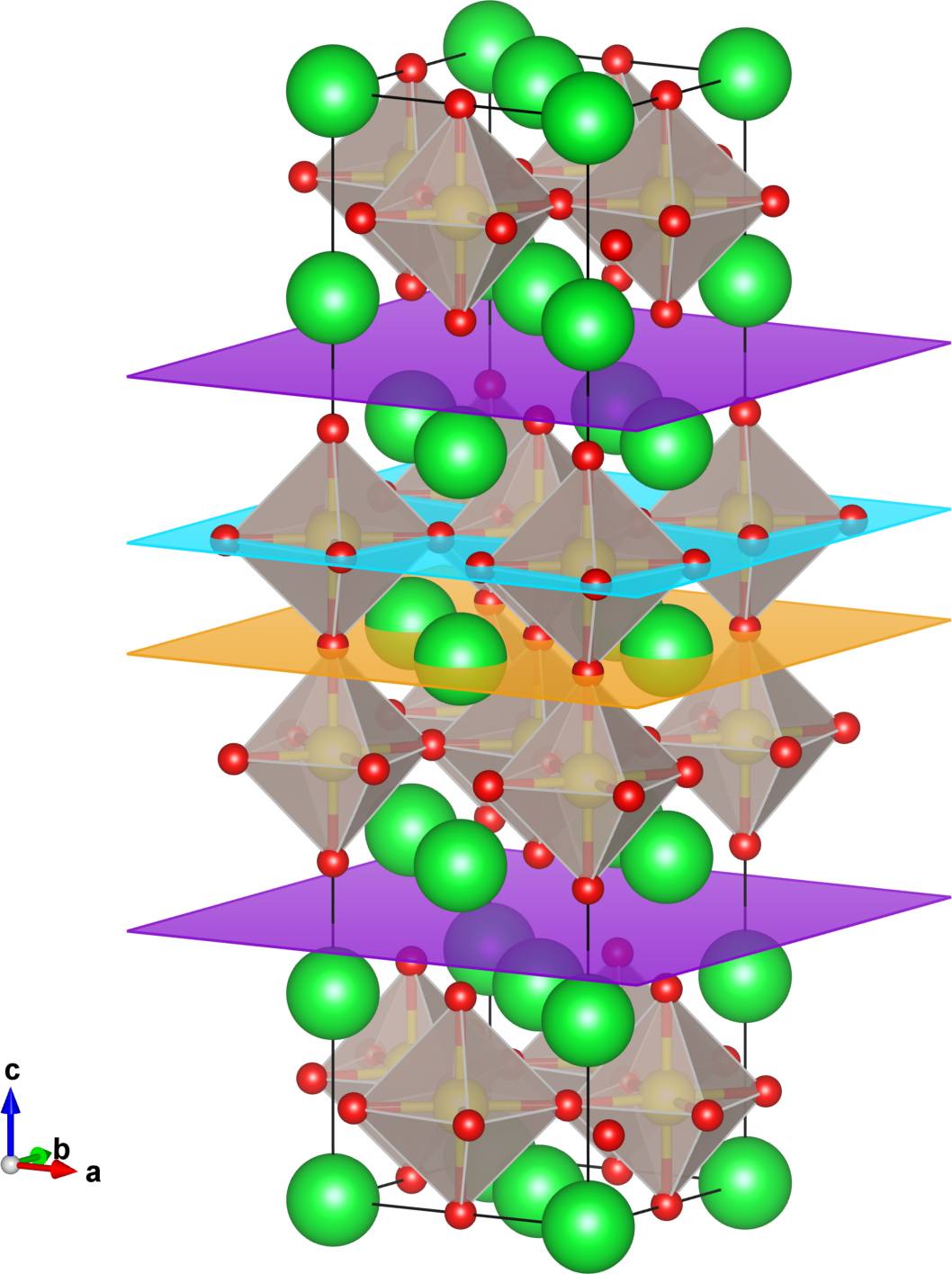}
\caption{Orthorhombic $Bbcb$  \sro~crystal structure with the calculated cleavage planes. The rock-salt like (violet) and SrO (orange) cutting planes are both SrO terminated, while the turquoise cleavage plane is RuO$_{2}$ terminated.
Sr: green, Ru: yellow, O: red.} 
\label{fig:sro-bulk}
\end{center}
\end{figure}

The \sro~crystal is the $n=2$ member of the Ruddelsden-Popper series Sr$_{n+1}$Ru$_{n}$O$_{3n+1}$. It consists of two perovskite  SrRuO$_3$ layers separated by a rock-salt like SrO terminated interface, as shown in Fig. \ref{fig:sro-bulk}. The RuO$_6$ octahedra are rotated by 6.8$^{\circ}$ around the [001] direction \cite{Shaked2000}, but in contrast to \cro, no tilting is observed. The connected octahedra within each double row are rotated against each other, resulting in an orthorhombic unit cell (space group $Bbcb$) with a square basal plane for \sro\ which  is referred to as \ct~structure in the present paper.

The optimized lattice constant of the orthorhombic bulk \sro~unit cell ($a=b= 3.899$\,\AA, $c= 20.722$\,\AA) was determined using the van-der-Waals (vdW) corrected \opt~functional. The calculated geometries are in very good agreement with experimental values \cite{Shaked2000} ($a=b= 3.890$\,\AA, $c= 20.725$\,\AA), showing deviations below 0.25\%. Compared to  experiment, the rotation of the octahedra is slightly larger, a consequence of the increased Ru--O bond lengths. The preferred cleaving plane is between two SrO layers at the rock-salt like interface with a calculated surface energy of 49\,meV/\AA$^2$, consistent with previous experiments \cite{Stoger2014a}. 

We have also calculated the surface energies of the (001) cleaving planes at the SrO and the RuO$_{2}$ interface, which yield surface energies of 76\,meV and 114\,meV, respectively. In both cases, the strong Ru--O bonds have to be broken, leading to these high values compared to the rock-salt like interface. It can be expected that other planes cutting through Ru--O bonds, like e.g. (100), would be similarly unfavorable.

\begin{figure}[htbp]
\begin{center}
\includegraphics[trim= 0cm 0cm 0cm 0cm, clip, width=\textwidth] {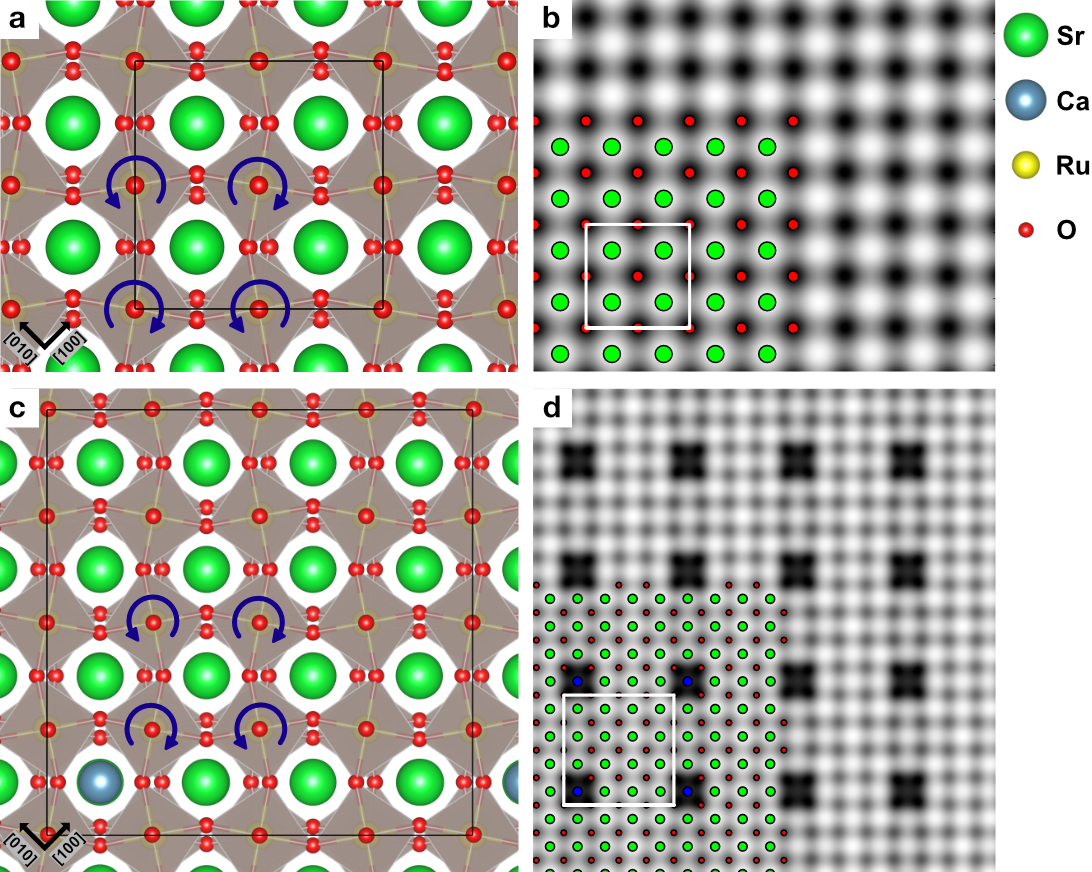}
\caption{(a):  \ct~\srosurf~surface model cell. (b):  corresponding STM simulation (\srosurf), +0.5\,V bias voltage. 
(c): \cf\ Ca-doped \srosurf~model cell (6.25\% dopant concentration). (d): corresponding STM simulation (Ca-doped \srosurf), +0.5\,V bias voltage.
\label{fig:sro-substrate}}
\end{center}
\end{figure}

At the surface, the rotation of the octahedra is slightly increased to 11$^{\circ}$. The Ca-doped surface was simulated by replacing a single surface Sr atom of a \cf\ unit cell by a Ca atom, yielding a dopant concentration of 6.25\%. The relaxation of the doped structure shows that the Ca dopant occupies the same high symmetry position of the original Sr atom, but is slightly pushed (0.17\,\AA) into the surface. The Ca dopant does not significantly change the DOS of the substrate (not shown), since the main difference to the Sr states is found in the position of the Ca 3$s$ and 3$p$ states at $\approx -20$\,eV while the depleted Ca 4$s$ states at the Fermi level are rather similar to the Sr 5$s$. Close to the Fermi edge, the substrate DOS is dominated by the Ru $d$ and O $p$ states, which play the primary role in the hybridization with the adsorbate. 
A Scanning Tunneling Microscopy (STM) image simulation of the pristine surface (Fig. \ref{fig:sro-substrate}) exhibits a perfectly regular grid of bright dots at the position of the surface Sr atoms at both positive and negative bias voltages. The Ca-doped surface on the other hand shows the Ca dopant as a dark feature in the bright Sr rows, well in line with experimental measurements \cite{Stoger2014a}.

\subsection{Low Coverage of \oom}
\begin{figure}[htbp]
\begin{center}
\includegraphics[trim= 0cm 0cm 0cm 0cm, clip, width=\textwidth]{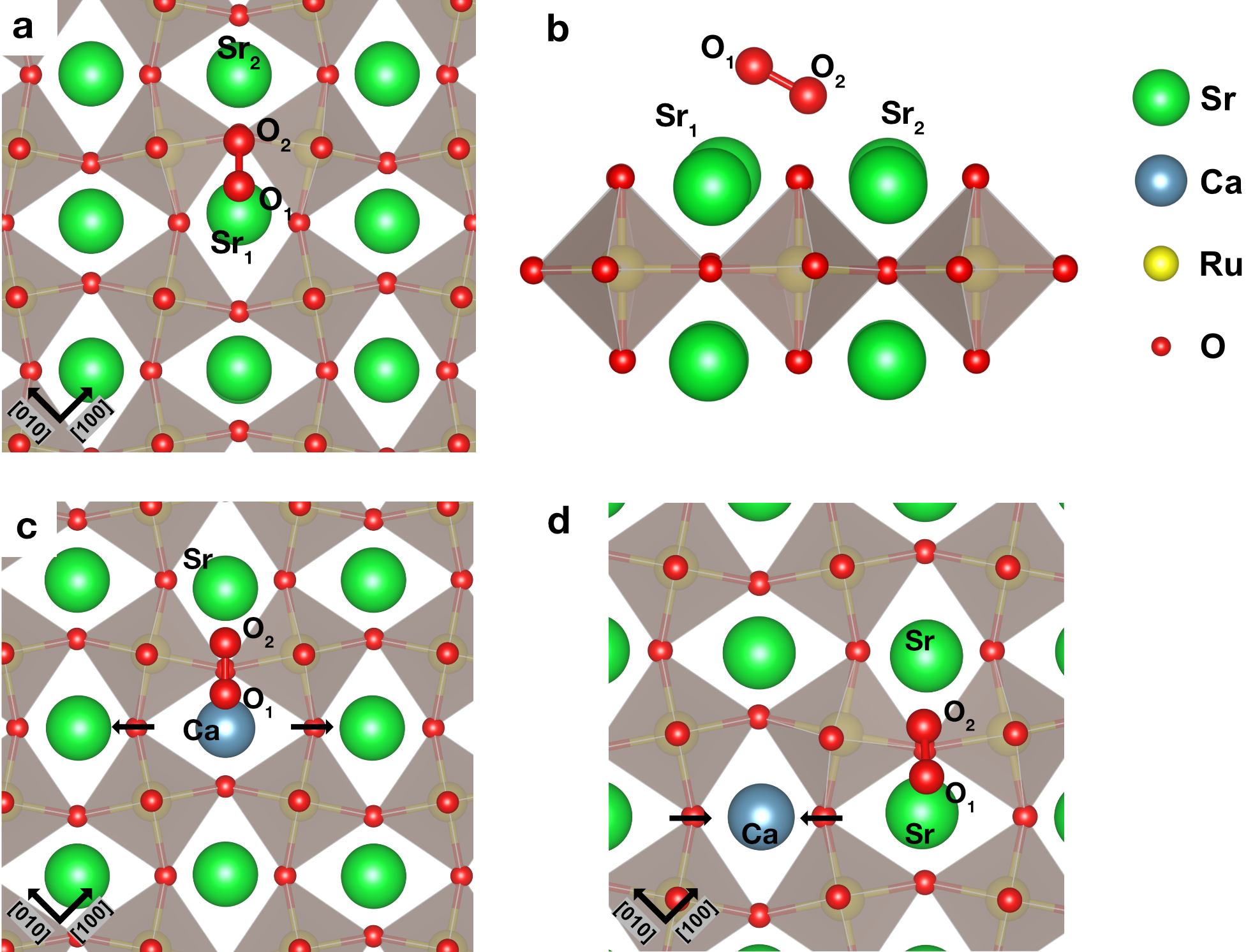}
\caption{Low coverage of \oom~adsorbed on (a)-(b) pure and (c)-(d) Ca-doped \srosurf. For Ca-doping, (c) shows the ``wide'' and (d) the most favored configuration.\label{fig:sro-lowcoverage}}
\end{center}
\end{figure}

In the low coverage case, a single \oo~molecule was adsorbed in the \cf\ model cell, yielding 1/16 monolayer (ML) coverage with respect to  the number of surface RuO$_6$ octahedra. It adsorbs as a charged superoxo \oom, indicated by an 12\% elongation of the O--O bond to 1.35\,\AA. In the optimized geometry, the molecule is tilted by 28.8$^{\circ}$ and located at a Sr--Sr bridge site, close to one of the neighboring Sr atoms (Figs. \ref{fig:sro-lowcoverage}(a,b)). Both distances d(O$_{1}$--Sr$_{1}$) and d(O$_{2}$--Sr$_{2}$) to the respective closest Sr atom is 2.51\,\AA. The octahedra at each side of the adsorbed molecule are tilted away 3$^{\circ}$ from the [001] axis. The vdW corrected \opt~functional clearly preferes a molecular adsorption ($\ea = -1.42$\,eV) compared to the dissociated molecule ($\ea = -0.67$\,eV).

On the Ca-doped surface the rotation of the octahedra in the first subsurface layer gives rise to two distinct adsorption configurations close to the Ca dopant. The rotation of the octahedra in the subsurface enables the Ca atom to move more easily in the direction where the octahedra are rotated away from it, giving rise to a ``wide'' (see Fig. \ref{fig:sro-lowcoverage}(c)) and a ``narrow'' configuration with respect to the orientation of the adsorbed molecule. 
In both cases, the \oom~molecule lies fairly flat (4.7$^{\circ}$ and 7.9$^{\circ}$, respectively) and the adsorption energies are very similar ($-1.43$\,eV and $-1.41$\,eV, respectively). Compared to the undoped case, the atomic distances d(O$_{1}$--Ca$_{1}$) and d(O$_{2}$--Sr$_{2}$) to the closest surface ions change from 2.51\,\AA\ to 2.38\,\AA\ and 2.60\,\AA. The Ca dopant is slightly displaced only in the ``narrow'' adsorption configuration. Interestingly, the most reactive adsorption site was found at a Sr--Sr bridge next to the dopant site (Fig. \ref{fig:sro-lowcoverage}(d), $\ea= -1.49$\,eV), most probably due to the resulting higher flexibility of the surface compared to the undoped case. Inspecting the DOS of an \oom\ adsorbed at an Ca--Sr or Sr--Sr bridge site, one finds that the \oom\ states shift upwards by about 0.3\,eV at Ca--Sr bridges compared to the Sr--Sr bridges, which explains the slightly preferred adsorption at the latter site. Nevertheless, the specific site does not influence the charge state of the adsorbed molecule.

\subsection{High coverage, from 1/4 monolayer to full coverage}
Higher coverages of \oom~molecules were only studied on the undoped system. At 1/4 monolayer (ML) coverage the \oom~adsorbs at a Sr--Sr bridge in an geometry similar to the low coverage case. Again, the \oom~is slightly displaced towards one of the Sr ions and is tilted upwards. Due to the charge transfer the oxygen bond length is again increased to  1.35\,\AA\ and the Sr--O distances are slightly more symmetric compared to the low coverage case at 2.49\,\AA\ and 2.52\,\AA\ for d(O$_{1}$-Sr$_{1}$) and d(O$_{2}$-Sr$_{2}$), respectively. The adsorption energy per molecule is only slightly reduced to $-1.40$\,eV (Tab. \ref{tbl:sro327-2l-1O2-eads}), indicating only a rather small adsorbate-adsorbate interaction for the 1/4\,ML coverage regime.

\begin{figure}[htbp]
\begin{center}
\includegraphics[trim= 0cm 0cm 0cm 0cm, clip, width=\textwidth]{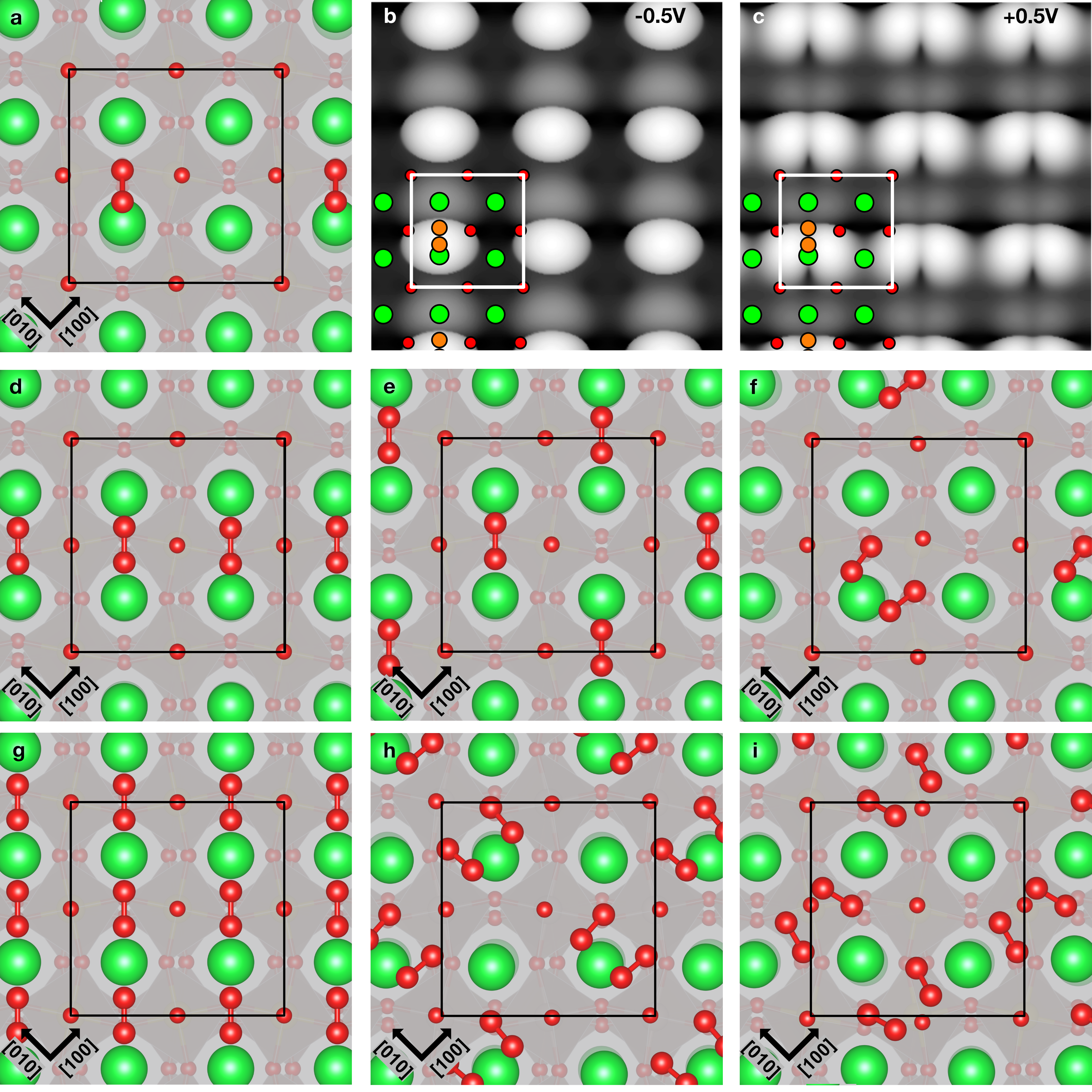}
\caption{High coverage structures. (a)-(c): 1/4\,ML coverage of \oom~located at a Sr--Sr bridge site. STM simulations show (b) a single bright feature at $-0.5$\,V bias voltage and (c) a double lobed bright feature at +0.5\,V bias voltage. The bright blob is separated along the directional axis of the \oom. (d)-(f): 1/2\,ML coverage, \oom~arranged in a parallel, zig-zag and clustered order, respectively. (g)-(i): full coverage structures for in the line, rotated clustered, and parallel clustered arrangements, respectively.\label{fig:sro-highcoverage}}
\end{center}
\end{figure}

Two typical simulated STM images are shown in Fig. \ref{fig:sro-highcoverage}(b) and  Fig. \ref{fig:sro-highcoverage}(c). The adsorbed molecules are imaged as a protrusion centered at the high O-atom of the tilted \oom. Upon adsorption, the charge transfer to the \oom\ molecule breaks the 2-fold degeneracy of  the triplet \tpps\ molecular orbital and splits the  orbital into a fully  occupied \tppsper\ orbital perpendicular to the surface and a half-filled \tppspar\ orbital parallel to the surface \cite{Halwidl2018}, where the latter is clearly visible in the simulations at positive bias voltage.

Upon increasing the \oom~coverage the adsorbed oxygen molecules start to experience repulsive interactions (see Tab. \ref{tbl:sro327-2l-1O2-eads}). For a coverage of 1/2\,ML (two \oo~molecules in a \ct~model cell) more complex adsorption configurations start to form. On the DFT level the fully relaxed structures show that neither the formation of parallel \oom\ lines along the [010] direction ($\ea= -0.98$\,eV, see Fig. \ref{fig:sro-highcoverage}(d)) nor along the [100] direction ($\ea= -1.00$\,eV) is  favored, as a zig-zag arrangement of the molecules  (Fig. \ref{fig:sro-highcoverage}(e)) leads to an increased adsorption energy per molecule of $\ea= -1.12$\,eV. In that case, every other bridge site is occupied and the rows are shifted by half a unit cell. Additionally, the potential energy landscape of adsorption is very shallow, allowing a multitude of adsorption configurations: also a rotation of up to 18$^{\circ}$ away from the Sr--Sr bridge orientation is energetically almost degenerated (< 10\,meV). This rotation is also coupled with an increase in the tilting angle from around 1$^{\circ}$ to a value of up to 27$^{\circ}$. 

Interestingly, the DFT calculations predict a clustering of the \oom\ molecules at a Sr atom in the most stable adsorption configuration with an adsorption energy per \oom\ of $\ea= -1.14$\,eV, shown in Fig. \ref{fig:sro-highcoverage}(f).  In this configuration, both \oom~molecules are tilted from the surface by 27$^{\circ}$ and share an angle of 14$^{\circ}$ between each other. 

The full coverage limit, i.e. one \oom~per RuO$_6$ octahedron, was investigated  with four oxygen molecules in the \ct~surface unit cell. Again, the repulsion between the \oom\ molecules makes a line-wise arrangement unfavorable (Fig. \ref{fig:sro-highcoverage}(g)). Instead, a low symmetry adsorption configuration is preferred where the \oom~molecules show a mixture of strongly and weakly tilted arrangements, grouped around a surface Sr atom. The oxygen molecules are displaced from the Sr--Sr bridge sites to a Sr--O bridge site between a surface ion and the apical O of a nearby RuO$_6$ octahedron, as shown in Figs. \ref{fig:sro-highcoverage}(h,i). 
Two clustered configurations were found:  In the most favored case, shown in Fig. \ref{fig:sro-highcoverage}(i) ($\ea= -0.95$\,eV), the oxygen molecules are clustered in groups of two around an apical oxygen atom, where one oxygen of each molecule is located at the centre of a Sr--Sr bridge, while the other oxygen is tilted upwards by 16$^{\circ}$ to 30$^{\circ}$ and rotated towards a Sr--O bridge. The octahedra around which the \oom~molecules are grouped tilt by 4$^{\circ}$ and 8$^{\circ}$. The two mentioned groups show a mirror symmetry along the ($1\overline{1}0$) plane, the molecules therefore form a slightly wavy double line along the [$\overline{1}10$] direction. The O--O length is slightly reduced to 1.28\,\AA. A similar arrangement, shown in Fig. \ref{fig:sro-highcoverage}(h), where the two groups are rotated by an angle of 90$^{\circ}$ towards each other yields an energy penalty of 20\,meV. In this arrangement all \oom~molecules are clustered around a surface Sr and tilt by 21$^{\circ}$ to 23$^{\circ}$. 

\begin{table}[htp]
\caption{\oo~adsorption energy per molecule on Ca-doped and pure \srosurf.}
\begin{center}
\begin{tabular}{l c c c}
\hline
				&\multicolumn{3}{c}{$\ea$~[eV/\oo]} 	 \\
				&{\opt}	&{\hse}	&{RPA}		\\
\hline
& \multicolumn{3}{c}{1/16\,ML coverage}	\\
Sr--Sr				&-1.42	&	&	\\
Sr--Sr, near Ca		&-1.49	&	&	\\
Ca--Sr, wide		&-1.43	&	&	\\
Ca--Sr, narrow		&-1.41	&	&	\\
\hline
& \multicolumn{3}{c}{1/4\,ML coverage}	\\
Tilted			&-1.40	&-0.91	&-0.99	\\
Dissociated		&-0.67	&-0.09	&-0.02	\\
\hline
& \multicolumn{3}{c}{1/2\,ML coverage}	\\
Parallel			&-0.98	&-0.68	&-0.82	\\
Zig-zag			&-1.11	&-0.75	&-0.85	\\
Clustered			&-1.14	&-0.29	&-0.52	\\
\hline
&\multicolumn{3}{c}{full 1\,ML coverage}	\\
Sr--Sr bridge, line	&-0.82	&-0.43	&-0.49	\\		
Clustered, parallel	&-0.95	&-0.07	&-0.33	\\
Clustered, rotated	&-0.92	&-0.01	&-0.25	\\
\hline
\end{tabular}
\end{center}
\label{tbl:sro327-2l-1O2-eads}
\end{table}

\subsection{RPA adsorption energies}
As shown in a recent publication \cite{Halwidl2018}, the vdW-DF functional strongly overestimates the adsorption energies of the superoxo species on \cro, in part due to the overestimation of the electron affinity of the \oo~molecule. 
We therefore evaluated potential error bars by performing many-electron (RPA) total energy calculations 
for both  the low and high coverage limit structures (1/4\,ML and full coverage, respectively) in the small \ct~unit cell. 
Already in the low-coverage case of 1/4\,ML our results show that the adsorption energy per \oom\ is reduced by $\sim$\,0.4\,eV on the RPA level (Tab. \ref{tbl:sro327-2l-1O2-eads}), nearly identical to the value found for \cro. 
Additionally, the RPA values for the coverage of 1/2\,ML indicate that the repulsion between the \oom\ molecules is underestimated on the DFT level. 
While the adsorption energy per \oom\ of the regular (parallel or zig-zag) structures is again reduced by $\sim$\,0.2-0.3\,eV  (Tab. \ref{tbl:sro327-2l-1O2-eads}), the most stable DFT configuration where the \oom\ are clustered at a surface cation is now strongly disfavored ($\ea= -0.52$\,eV). 
In addition to the much smaller geometric distance between the \oom\ molecules, an additional contribution to the enhanced repulsion can tentatively be assigned to the different hybridization of the clustered \oom\ molecules with the surface. In contrast to the regular (line-wise) adsorption configurations, where the additional electron is transferred to the originally unoccupied \oo\ \tppspar\ orbital (see discussion below), the additional tilting of the \oom\ molecule in the cluster configuration results in a rehybridization where the \oo\  \tppsper\ orbital is also partially emptied. This fractional occupation is already strongly penalized on the HSE level, and a similar effect is observed on the RPA level (see Tab. \ref{tbl:sro327-2l-1O2-eads}).

This trend continues at full coverage. In contrast to the  DFT results, where the adsorption energy per \oom\ of the  favored cluster configuration is just about 0.19\,eV lower compared to the most favored 1/2\,ML coverage structure, the RPA results show enhanced repulsion for this configuration. Consequently, on the RPA level the most stable configuration is the double line structure with an adsorption energy of 0.49\,meV per \oom.

\begin{figure}[htbp]
\begin{center}
\includegraphics[trim= 0cm 0cm 0cm 0cm, clip, width=\textwidth]{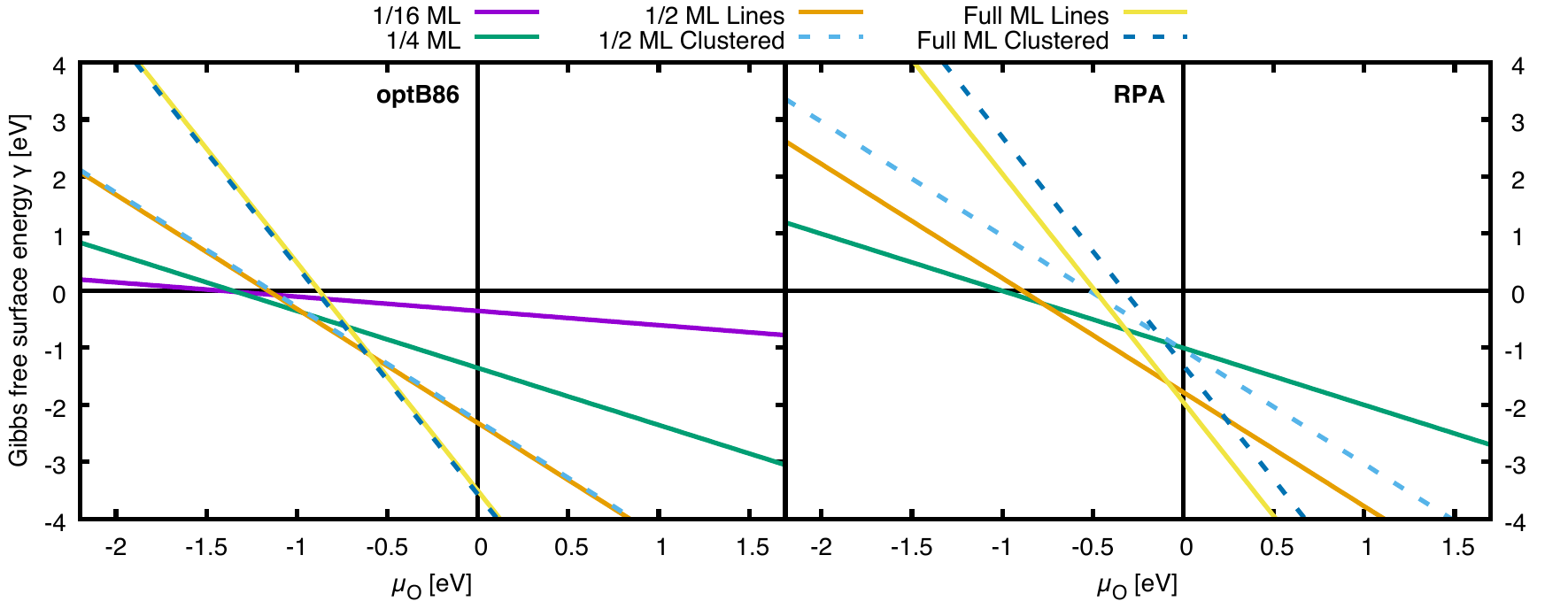}
\caption{Phase diagrams of adsorption calculated with the \opt~vdW-DFT functional and the many-electron RPA method as a function of the chemical potential $\mu_O$ of oxygen. The correspondence between the labelled lines and the adsorption configurations is as follows: 1/16\,ML $\to$ Fig. \ref{fig:sro-lowcoverage}(a),  1/4\,ML $\to$ Fig. \ref{fig:sro-highcoverage}(a),
1/2\,ML Lines $\to$ Fig. \ref{fig:sro-highcoverage}(d), 1/2\,ML Clustered $\to$ Fig. \ref{fig:sro-highcoverage}(f), 
Full\,ML Lines $\to$ Fig. \ref{fig:sro-highcoverage}(g), Full\,ML Clustered $\to$ Fig. \ref{fig:sro-highcoverage}(i).
\label{fig:sro-o2-phasediagram}}
\end{center}
\end{figure}

The rather large differences between the DFT  (\opt) and RPA results are also clearly reflected in a predicted phase diagram for \oo~adsorption (see Fig. \ref{fig:sro-o2-phasediagram}). 
The DFT results suggest that a full monolayer can be reached under typical ultra-high vacuum conditions of $10^{-11}$\,mbar and liquid nitrogen temperatures, which corresponds to an oxygen chemical potential of $\mu_{O}=-0.15$\,eV. 
DFT also shows clustered and line configurations to be equally stable. However, the phase diagram based on the RPA values makes a full monolayer unlikely and clearly favors evenly spaced adsorbate configurations.

\subsection{Oxygen Charge State}
\begin{figure}[htbp]
\begin{center}
\includegraphics[trim= 0cm 0cm 0cm 0cm, clip, width=0.4\textwidth]{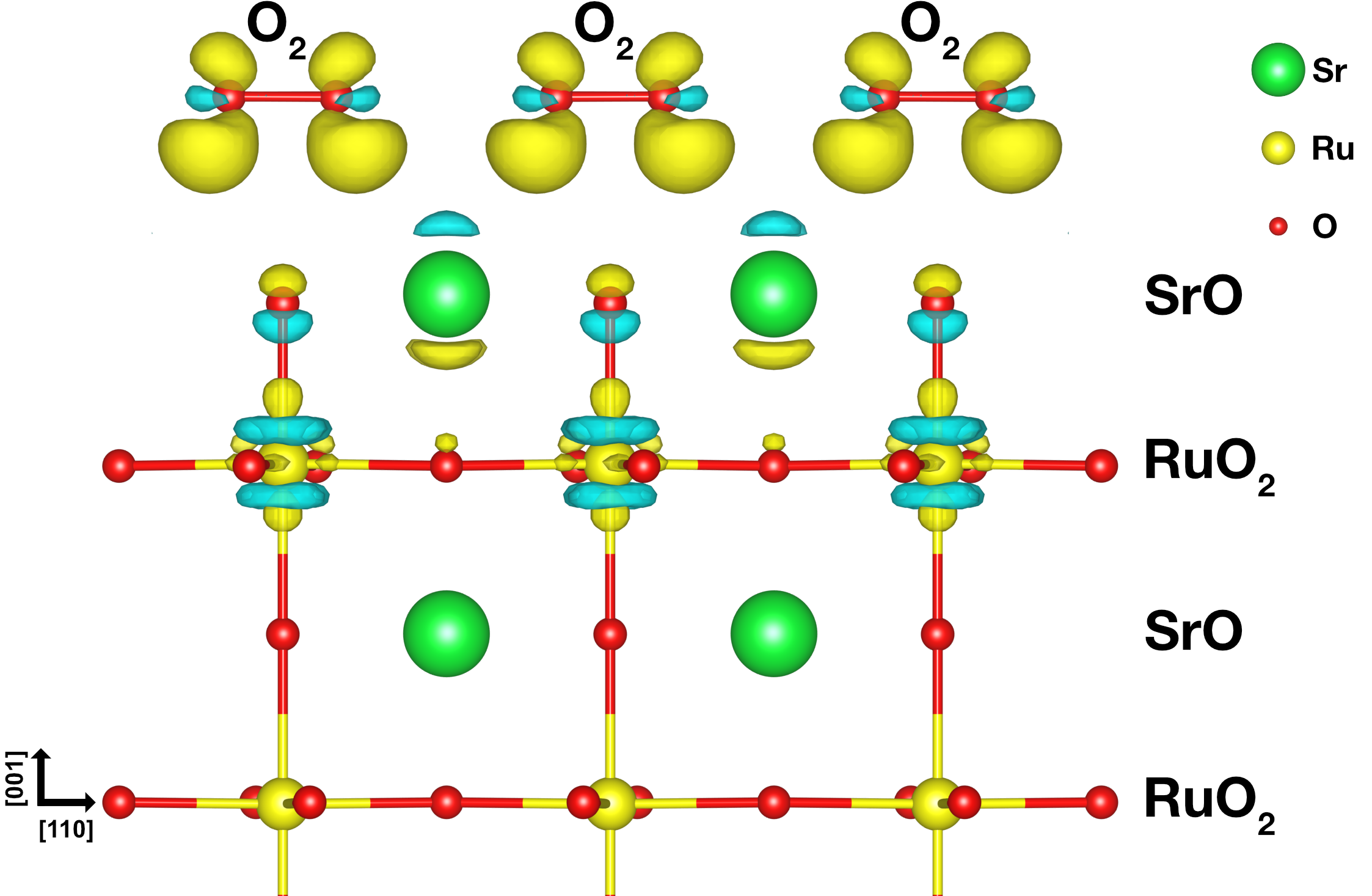}
\caption{Charge density difference plot at full \oo~coverage calculated with the \opt~functional. Yellow and turquoise areas indicate charged and depleted areas, respectively.
\label{fig:sro-o2-chgdiff}}
\end{center}
\end{figure}

\begin{figure}[htbp]
\begin{center}
\includegraphics[trim= 0cm 0cm 0cm 0cm, clip, width=\textwidth]{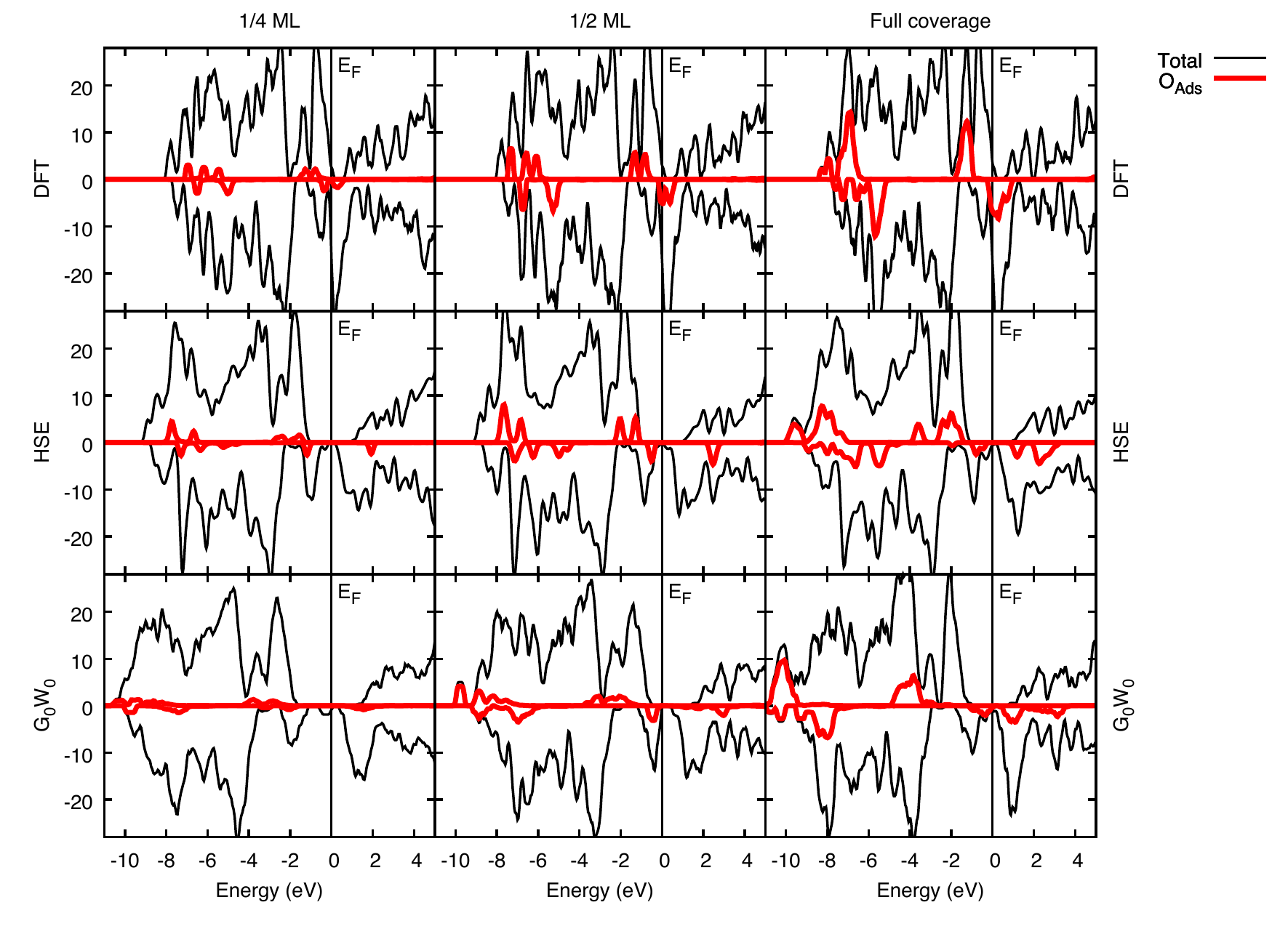}
\caption{Density of states (DOS) for 1/4\,ML, 1/2\,ML, and full ML \oo~coverage calculated with the \opt~and \hse~functional, and the post-DFT many-electron \gw~method.
\label{fig:sro-o2-dos}}
\end{center}
\end{figure}

The calculated spin-polarized Density-of-States (DOS) of the adsorbed oxygen molecule displayed in Fig. \ref{fig:sro-o2-dos} reveals a charge transfer to the adsorbed \oo, which creates a charged superoxo \oom~species at all coverages on both bare and Ca-doped \srosurf. The originally half-filled \oo-\tpps\ molecular orbital has its two-fold degeneracy lifted upon adsorption and the charge is transferred to the lower lying empty \oo-\tppsper\ spin-down channel. Particularly at low coverage, the now fully occupied \oom-\tppsper\ molecular orbital points towards the surface Sr ions. As shown for the high coverage limit in Fig. \ref{fig:sro-o2-chgdiff}, the charge originates from the RuO$_{2}$ layer of the substrate while the surface Sr atoms close to the adsorption site do not contribute at all, as we have shown for the \oo/\crosurf\ system \cite{Halwidl2017}. The Sr and O ions at the interface close to the adsorbate are just slightly polarized. An analysis of Bader charges \cite{Henkelman2006,Steffen2010,Tang2009} shows that the Ca and Sr formal charge of +2 is unchanged at all coverages. The evaluation of the vibrational frequencies at 1/4\,ML coverage shows a red shift to 1132\,cm$^{\mbox{-}1}$ of the stretching mode of the \oom, which fits well to the experimental value of a charged \oom~in the gas phase \cite{Krupenie2012}. The calculations do not yield an unstable mode. 
 
However, a closer inspection of the DOS also reveals the shortcomings of a DFT description of the charged \oo~molecule. As shown in Fig. \ref{fig:sro-o2-dos}, the HOMO-LUMO gap of the \oom~molecule as calculated by DFT is too small (0.5\,eV) and as a consequence the electron affinity of an adsorbed \oo~molecule is overestimated, i.e. the cost to transfer charge is decreased. As a result a too large adsorption energy per \oom, which is referenced to the neutral \oo\ molecule, is predicted. This issue can be alleviated by using a hybrid DFT functional like \hse~or a post-DFT many-electron ansatz like \gw. 
For 1/4\,ML coverage, the HOMO-LUMO gap of the \oom~molecule is increased to 3.1\,eV and the predicted adsorption energy per \oom\ is reduced to $-0.59$\,eV. Using the many-electron \gw~approximation the HOMO-LUMO gap is increased to 3.4\,eV and the corresponding RPA adsorption energy per \oom\ is $-0.99$\,eV. This trend continues for 1/2\,ML coverage, here the \hse~and \gw~HOMO-LUMO gap of the adsorbed \oom~is increased to 2.9\,eV and 3\,eV, respectively. 
At full coverage for the most stable structure, the double line adsorption configuration depicted in Fig. \ref{fig:sro-highcoverage}(g), the increase of the HOMO-LUMO gap over the DFT estimate is less pronounced for both \hse~and \gw, presumably due to the broadening of the \oom~states, caused by an increased adsorbate-adsorbate interaction. 
While the DFT DOS for the higher coverage structures might suggest coexistence of differently charged oxygen molecules, the examination of the DOS of the individual molecules shows that all contain the same amount of charge, identified as superoxo \oom. 

Further charging of the \oom\ molecule yielding a peroxo species only occurs in the presence of additionally available electrons. In contrast to the adsorption of \oo\ on La$_2$NiO$_4$ where the electron is supplied by the surface cation \cite{Akbay2016}, the Bader charge density analysis for the present case shows that the formal charge on either Ca or Sr is unchanged from the +2 on the clean surface. However, by artificially charging the unit cell at 1/4\,ML coverage a peroxo species is formed at the DFT level of theory with an energy penalty of about 0.4\,eV, which roughly corresponds to the energy of the unoccupied \tpps\ orbital of the superoxo species with respect to the substrate Fermi level. However, as mentioned before, this penalty is again strongly biased by the too small HOMO-LUMO gap of the \oom\ superoxide. Therefore, at the more appropriate many-electron (\gw) level of theory, which yields a larger HOMO-LUMO gap, the stability of a peroxo species is expected to be much lower.

\section{Discussion}

According to our recent findings \cite{Halwidl2018}, \oo~adsorbs as a superoxo \oom~species on the \crosurf~surface at all coverages. 
Although  \sro~is closely related to \cro,  there is a structural difference due to the additional tilting of the RuO$_6$ octahedra in \cro. 
This tilting causes a channel-like pattern of wider and narrower surface oxygen rows at the CaO terminated \crosurf~surface, 
while the symmetry of the SrO terminated \srosurf\ surface layer is identical to square symmetry of a binary rock-salt SrO(001), albeit at an expanded \sro~lattice constant. 
One way to disentangle the structural effects from the chemical modifications due to the different cation termination is to study Ca-doped \srosurf. 
On both surfaces the \oo~molecules preferentially adsorb at or near a cation-cation bridge as a charged superoxo \oom~molecule.

To accommodate for the optimal atomic distance to the \sro\ surface cation, the \oom~shifts towards one of the surface Sr, and tilts upwards at the same angle, 29$^{\circ}$, as on \crosurf. 
Inserting a Ca dopant in the \srosurf\ surface layer results in a nearly identical adsorption energy at the Ca--Sr site. Nevertheless, a moderate increase of the adsorption energy by $\sim$\,80\,meV is found for the Sr--Sr site next to the Ca dopant. 
At higher coverages, the structural differences of the adsorption configurations on \sro\ and \cro\ subside. Like on \crosurf, at 1/2\,ML coverage the adsorbed \oom~molecules start to rotate away from the highly symmetric Sr--Sr bridge sites towards the Sr--O bridge positions between a surface ion and the apical oxygen of a nearby octahedron. 
Our calculations show a slightly higher adsorption energy on \sro\ compared to \cro.
However, as observed for \crosurf\ \cite{Halwidl2018}, the adsorption energy of an \oo~molecule on \srosurf\ is significantly overestimated, partially due to the overestimation of the \oo\ electron affinity. 
A comparison to the values derived from many-electron (RPA) calculations  shows a decrease of about 0.4\,eV for both  \crosurf\ and \srosurf. 

The most striking difference between the adsorption of \oo\ on  \sro\ and \cro\ is related to surface structure: While the channel structure of the \cro\ surface strongly restricts the accessible 
surface sites, the high symmetry of the \sro\ surface layer leads to flat DFT potential energy surface for the adsorption, and consequently a higher number of energetically nearly degenerated configurations at increased 
\oom\ coverage. These configurations include \oom\ clusters with a fractional occupation of the O$_2$  \tpps\ orbitals.  
However, the RPA calculations reveal that the clustered configurations are an artifact of the (vdW) DFT, and yield higher stability for adsorption at Sr--Sr bridges, in line with the findings for \oom~on \crosurf.  Furthermore the RPA surface \oo~phase diagram clearly predicts that higher \oom~coverages than 1/2\,ML are quite unlikely on \srosurf, which is in accord with the findings on \crosurf~where a saturation coverage of 1\,ML was found, corresponding to a 1/2\,ML coverage on \srosurf.

\section{Conclusion}
To summarize, the main adsorption mechanism of \oo\ on \srosurf\ is very similar to \crosurf\ \cite{Halwidl2018}. This concerns both the preferred adsorption sites at or near Sr--Sr bridges and the charging of the adsorbed \oom~molecule by a charge transfer from the subsurface RuO$_2$ layer, without the necessity of dopants or surface defects. The main difference is related to the higher symmetry of the \sro\ surface, which allows for an increased number of adsorption configurations.
Yet our results show a pronounced shortcoming in the DFT description of the electronic structure, resulting in an overestimation of the adsorption energies. The comparison with many-electron approaches (\gw\ and RPA) 
suggests an overestimation of the vdW-DF adsorption energies about 0.4\,eV, resulting in values of   $-0.99$\,eV and $-0.49$\,eV for low and high coverages respectively. Additionally, they strongly disfavor the formation
of local \oom\ clusters predicted by standard DFT functionals.

\section*{Acknowledgments}
This work was supported by the Austrian Science Fund (FWF project F45 ``FOXSI'') and the Vienna Scientific Cluster (VSC). The structural models were created with the program VESTA \cite{Momma:db5098}.


\end{document}